\documentclass[%
 preprint, 
 amsmath,amssymb,
 aps, physrev,
prd,
]{revtex4-2}
\usepackage{graphicx}
\usepackage{dcolumn}
\usepackage{bm}
\usepackage{hyperref}

\begin{document}


\title{\textbf{General Formulation of Covariant Tensor Amplitudes for Purely Hadronic and Single-photon Radiative Cascade Decays} 
}%

\author{Yu-chang Sun}
\author{Yu-xuan Hu}
\author{Chu-cheng Pan}
\author{Hao Cai}%
 \email{Contact author: hcai@whu.edu.cn}
\author{Liang Sun}%
 \email{Contact author: sunl@whu.edu.cn}
\affiliation{%
Wuhan University, School of Physics, Wuhan, Hubei, China.
}%

\date{\today}

\begin{abstract}
We present a unified method for generating Lorentz-covariant partial-wave amplitudes for multi-body purely hadronic and single-photon radiative cascade decays within the covariant $L$-$S$ tensor formalism. Based on the structural decoupling of external polarization wave functions, complex coupling parameters, and the core covariant amplitude tensor $U$, we establish a hierarchical vertex contraction procedure that explicitly incorporates intermediate-state cascade bridge projection operators $P^{(J_i)}(p_i)$ and parent-frame definite-spin projection operators $P^{(S_{BC})}(p_A)$. 
This rigorous treatment guarantees the Rarita-Schwinger symmetry, tracelessness, and transversality conditions across different reference frames, fundamentally eliminating the unphysical longitudinal and trace contaminations induced by Lorentz boosts in cascade decays. 
Furthermore, we formulate a general covariant tensor representation for single-photon radiative decays $A(J_M) \to \gamma + X(J_X)$ that strictly satisfies photon gauge invariance ($q^\lambda V_{\dots\lambda\dots} = 0$) in the Lorenz gauge. 
Built upon normal- and anomalous-parity fundamental transverse tensors ($K_{\mu\nu}$ and $S_{\mu\nu}$), universal recursive formulas are derived via the systematic replacement $g_{\mu\nu} \to q_\mu p_{A\nu}$, accompanied by exact rules for matching the orbital angular momentum ranks of Blatt-Weisskopf centrifugal barrier factors $B_{L_{\text{rad}}}$ and a spectator index pass-through mechanism for the photon polarization index.
\end{abstract}

\maketitle


\section{\label{sec:level1}Introduction}

Hadron spectroscopy serves as a vital avenue for exploring the color confinement mechanism and the core dynamics of strong interactions within the non-perturbative regime of Quantum Chromodynamics (QCD). 
In recent years, with the stable operation of large-scale high-energy physics facilities such as BESIII, LHCb, Belle II, and GlueX, massive high-statistics datasets of hadronic and radiative decay events have been accumulated (for instance, the BESIII experiment has collected on the order of $10^{10}$ $J/\psi$ events and billions of $\psi(3686)$ events~[\onlinecite{Ablikim_2020}]). 
Searching for exotic hadronic states—such as glueballs~[\onlinecite{PhysRevLett.132.181901}], hybrids~[\onlinecite{PhysRevLett.129.192002}], and multiquarks~[\onlinecite{PhysRevLett.115.072001}]—within these multi-body final states, as well as precisely determining the masses, widths, and spin-parity quantum numbers ($J^{PC}$) of conventional hadron resonances, relies heavily on Partial Wave Analysis (PWA) techniques. 
By fully exploiting the invariant mass spectra and multi-dimensional angular distributions carried by the four-momenta of final-state particles, PWA decomposes the total scattering or decay amplitude into a coherent superposition of individual partial-wave amplitudes with definite quantum numbers. 
This enables the disentanglement and identification of the physical contributions from various intermediate resonances amidst complex interference patterns.   

Throughout the theoretical development of PWA, two mainstream representations for amplitude construction have been established: the Helicity Formalism~[\onlinecite{Jacob1959}] and the Covariant Tensor Formalism. 
Among them, the covariant tensor formalism aims to directly construct Lorentz-covariant amplitudes with definite $(L, S)$ quantum numbers. 
Originating from the non-relativistic three-dimensional Zemach tensor method~[\onlinecite{Zemach1965}], this formalism was subsequently combined with the symmetry, tracelessness, and transversality conditions of Rarita-Schwinger high-spin wave functions~[\onlinecite{Rarita1941}], and systematically generalized into a relativistic covariant form by Chung~\cite{Chung1993,Chung1998}, Filippini et al.~[\onlinecite{PhysRevD.51.2247}], and Zou and Bugg~[\onlinecite{Zou2003}]. 
In the covariant $L$-$S$ tensor scheme, scattering or decay amplitudes are directly constructed by contracting particle four-momenta, the Minkowski metric tensor $g_{\mu\nu}$, the totally antisymmetric Levi-Civita tensor $\epsilon_{\mu\nu\alpha\beta}$, and high-rank spin projection operators. 
This representation naturally satisfies Lorentz invariance, allowing direct computation using four-momenta in any reference frame without tedious Euler angle alignment rotations, while effortlessly incorporating the Blatt-Weisskopf barrier factors $B_L$ that describe centrifugal barrier suppression effects~[\onlinecite{vonHippel1972}]. 
Recently, theoretical investigations into covariant orbital-spin coupling schemes for arbitrary-spin particles based on irreducible tensors~\cite{Jing2023,Jing2024}, as well as covariant canonical-spinor amplitudes incorporating spinor-helicity variables~[\onlinecite{Huang2026}], have continued to advance, further expanding the theoretical foundation of covariant PWA. 

Meanwhile, the exponential growth in experimental statistics and the increasing topological complexity of cascade decays have placed stringent demands on both the automation level and computational performance of PWA software. 
Various computational tools for PWA have been developed internationally, such as the early GPUPWA framework~[\onlinecite{Berger2010}] widely used in the BESIII experiment to accelerate covariant tensor calculations via Graphics Processing Unit (GPU) parallelization, and the more recent TF-PWA framework~[\onlinecite{Jiang2024}], which is based on the helicity formalism and utilizes TensorFlow's Automatic Differentiation (AD) technology for generalized configurations. 
In large-scale unbinned maximum likelihood fits, the repeated iterative evaluation of the likelihood function and its gradients constitutes the primary computational bottleneck. 
In fact, regardless of whether the process involves a radiative photon, a complete covariant partial-wave amplitude can be strictly factorized mathematically into three mutually independent components: the polarization wave functions of external initial- and final-state particles $\varepsilon$ (where high-spin wave functions can be constructed by coupling fundamental wave functions via Clebsch-Gordan coefficients~[\onlinecite{Zhu1999}]), complex coupling parameters encapsulating resonance dynamical line shapes and relative strengths $\Lambda$, and the covariant amplitude tensor $U$ that purely describes the geometric coupling structure of orbital angular momentum and spin: 

\begin{equation}
A = \varepsilon_{\mu_1\cdots\mu_n}\sum_i\Lambda_iU^{\mu_1\cdots\mu_n}
\end{equation}

The recently proposed Partial Wave Analysis Code Generator (PWACG)~[\onlinecite{Dong2026}] fully demonstrates the core value of this structural decoupling in high-performance computing: because the covariant amplitude tensor $U$ for given $(L, S)$ quantum numbers depends solely on the four-momenta of experimental events and phase-space Monte Carlo (MC) samples—containing no free fitting parameters—it can be pre-computed and statically cached in GPU memory prior to the fitting procedure.
This "pre-computation and tensor caching" mechanism avoids redundant kinematic tensor evaluations during fitting iterations, making it feasible to integrate automatic differentiation with high-order Hessian matrix optimization algorithms such as the Newton-Conjugate Gradient (Newton-CG) method, thereby achieving orders-of-magnitude computational acceleration. 
However, to generalize this pre-computation architecture based on the covariant tensor $U$ into a universal framework supporting arbitrary topologies and arbitrary high spins, existing theories of covariant tensor construction still face two critical challenges:

\begin{enumerate}
    \item In multi-body purely hadronic cascade decays, the tensors used to construct decay amplitudes of secondary resonances satisfy transversality and tracelessness within their respective subsystem rest frames. 
    When these subsystem tensors enter the parent particle's rest frame for spin coupling, Lorentz boost effects cause the spin tensors—originally purely spatial in the subsystem frame—to acquire time-like (longitudinal) components and non-zero parent-frame trace components from the perspective of the parent frame, introducing unphysical degrees of freedom. 
    Following certain simplified treatments in traditional literature (such as the $(L, S) = (0, 1)$ and $(2, 1)$ partial-wave expressions given for $\psi \to \phi f_2$ in Ref.~[\onlinecite{Zou2003}]), performing naive four-dimensional metric contractions directly on subsystem analyzers without applying definite-spin projection in the parent frame allows unphysical longitudinal and trace contaminations to mix directly into the observable transverse amplitudes of the parent frame, causing the resulting amplitudes to deviate from the pure $L$-$S$ coupling subspace.
    \item In radiative cascade decays such as $J/\psi \to \gamma X$, the real photon, as a massless spin-1 gauge boson, possesses only two transverse polarization degrees of freedom. 
    Traditional tensor constructions for radiative decays typically provide specific expressions only for low-spin cases where the parent particle has spin 0 or 1~\cite{Zou2003,Dulat2005}.
    There is a lack of a general recursive formalism that strictly satisfies photon transversality (gauge invariance, $q^\lambda V_{\dots\lambda\dots} = 0$) in the Lorenz gauge, uniformly encompasses both normal- and anomalous-parity structures, and can be universally extrapolated to arbitrary parent and recoil particle spins $(J_M, J_X)$ while precisely matching the radiative orbital angular momentum barrier factors $B_L$. 
\end{enumerate}

To address these challenges, this paper develops a unified general formulation of covariant tensor amplitudes for both purely hadronic and single-photon radiative cascade decays. 
By explicitly incorporating intermediate-state cascade bridge projections and parent-frame definite-spin projection operators, we rigorously eliminate the unphysical longitudinal and trace contaminations induced by Lorentz boosts across reference frames. 
Furthermore, we establish universal recursive formulas for arbitrary-spin radiative decay vertices that strictly preserve photon gauge invariance, together with a spectator index pass-through mechanism for multi-level vertex contractions.

\section{Formalism for Covariant Amplitude Construction}

In partial wave analysis, after factoring out the polarization wave functions of external initial- and final-state particles as well as the complex coupling coefficients to be fitted, the core of amplitude evaluation reduces to constructing the covariant amplitude tensor $U$ with definite orbital angular momentum and spin-coupling quantum numbers. 
For a multi-body cascade decay process, the entire decay chain can be abstracted as a directed binary tree, where each internal node corresponds to a two-body decay vertex $A \to B + C$. 
To achieve the general construction of covariant amplitudes for arbitrary spins and cascade topologies, we establish a unified set of tensor construction and index contraction rules for both purely hadronic and single-photon radiative decay vertices below.

\subsection{General Form of Purely Hadronic Cascade Decay Vertices}
Consider a general hadronic two-body decay process $A(J_A, P_A; p_A) \to B(J_B, P_B; p_B) + C(J_C, P_C; p_C)$, where $J_i, P_i$, and $p_i$ ($i = A, B, C$) denote the spin, intrinsic parity, and four-momentum of the corresponding particle, respectively. 
According to the angular momentum coupling theorem, the spins of the final-state particles $B$ and $C$ first couple to form the total spin $S_{BC}$, which can take values in the range $\vert{}J_B - J_C\vert{} \le S_{BC} \le J_B + J_C$. 
Subsequently, $S_{BC}$ couples with the relative orbital angular momentum $L_{BC}$ between the two-body system to form the parent particle spin $J_A$, satisfying $\vert{}J_A - S_{BC}\vert{} \le L_{BC} \le J_A + S_{BC}$. 
In strong or electromagnetic interaction processes, if parity is conserved at the decay vertex, the allowed $(L_{BC}, S_{BC})$ combinations are further constrained by the parity selection rule: 
\begin{equation}
    P_A = P_B P_C (-1)^{L_{BC}}
\end{equation}
Within the covariant $L$-$S$ tensor formalism, the overall covariant amplitude for a single two-body decay vertex can be generally expressed as the Lorentz contraction of the parent particle polarization tensor $\epsilon_{(J_A)}(p_A, \lambda_A)$, the covariant parity factor $\tilde{E}(j_a, j_b, j_c)$, the orbital angular momentum tensor $T(L_{BC}; p_B, p_C)$, and the final-state spin-coupling polarization tensor $\Omega(S_{BC}; J_B, J_C)$: 
\begin{widetext}
\begin{equation}
    U_{A\to BC}(L_{BC}, S_{BC}; J_A, J_B, J_C) = \epsilon_{(J_A)}(p_A, \lambda_A) \, \tilde{E}(J_A, L_{BC}, S_{BC}) \, T(L_{BC}; p_B, p_C) \, \Omega(S_{BC}; J_B, J_C)
\end{equation}
\end{widetext}

The first term $\epsilon_{(J_A)}(p_A, \lambda_A)$ is the polarization tensor of the parent particle with spin projection $\lambda_A$. In practical calculations, for the primary decay vertex, this term stands independently as the first factor in the decoupled total amplitude and does not participate in the evaluation of the covariant amplitude tensor discussed in this paper. 
For a secondary decay vertex, this term is summed over spin projections together with the conjugate polarization tensor of the corresponding particle from the fourth term of the preceding vertex, thereby bridging the two consecutive decay vertices.  

The second term is the covariance factor $\tilde{E}(j_a, j_b, j_c)$. To preserve Lorentz covariance and parity transformation properties of the amplitude in four-dimensional Minkowski spacetime, this covariant factor is introduced according to the parity (even or odd) of the sum of the three coupled angular momenta, $j_a + j_b + j_c$: 
\begin{equation}
    \tilde{E}(j_a, j_b, j_c) = \begin{cases} 1, & j_a + j_b + j_c =\text{ even} \\ \epsilon_{\alpha\beta\gamma\delta} p_A^\delta, & j_a + j_b + j_c =\text{ odd} \end{cases}
\end{equation}
where $\epsilon_{\alpha\beta\gamma\delta}$ is the four-dimensional totally antisymmetric Levi-Civita tensor. When any two covariant tensors of ranks $r_1$ and $r_2$ couple to a target rank $r_{\text{target}}$, their index contraction follows a unified algebraic rule:
\begin{enumerate}
    \item When $r_1 + r_2 + r_{\text{target}}$ is even: The number of index pairs $n_c$ directly contracted between the two input tensors via the Minkowski metric $g^{\mu\nu}$ is given by: 
    \begin{equation}
        n_c = \frac{r_1 + r_2 - r_{\text{target}}}{2}
    \end{equation}
    \item When $r_1 + r_2 + r_{\text{target}}$ is odd: The covariant factor $\epsilon_{\alpha\beta\gamma\delta}p_A^\alpha$ is introduced, where the parent particle four-momentum $p_A^\alpha$ occupies the first index, and the two input tensors (requiring $r_1, r_2 > 0$) each contribute one spin index to contract with $\beta$ and $\gamma$, leaving $\delta$ as a free output index. The number of remaining index pairs $n_c$ contracted via the metric $g^{\mu\nu}$ is:
    \begin{equation}
        n_c = \frac{r_1 + r_2 - 1 - r_{\text{target}}}{2}
    \end{equation}
\end{enumerate}

The third term is the orbital angular momentum tensor $T(L_{BC}; p_B, p_C)$. The relativistically covariant orbital angular momentum tensor $\tilde{t}^{(L)}_{\mu_1 \cdots \mu_L}$ must satisfy the Rarita-Schwinger conditions: it is symmetric under the exchange of any two Lorentz indices, traceless upon contraction with the metric ($g^{\mu_i\mu_j}\tilde{t}_{\mu_1 \cdots \mu_i \cdots \mu_j \cdots \mu_L} = 0$), and orthogonal to the parent particle four-momentum ($p_A^{\mu_i}\tilde{t}_{\mu_1 \cdots \mu_i \cdots \mu_L} = 0$). 
This tensor is constructed by applying the rank-$L$ spin projection operator $P^{(L)}(p_A)$ to the direct product of the final-state relative four-momentum $r = p_B - p_C$, multiplied by the Blatt-Weisskopf barrier factor $B_L(Q_{abc})$ that describes centrifugal barrier suppression effects~[\onlinecite{Zou2003}]:
\begin{widetext}
\begin{equation}
    \tilde{t}^{(L)}_{\mu_1\cdots\mu_L}(p_A, p_B, p_C) = (-1)^L P^{(L)}_{\mu_1\cdots\mu_L \mu'_1\cdots\mu'_L}(p_A) r^{\mu'_1}\cdots r^{\mu'_L} B_L(Q_{abc}) 
\end{equation}
\end{widetext}
where $Q_{abc} = \sqrt{\frac{(s_A + s_B - s_C)^2}{4s_A} - s_B}$ is the magnitude of the three-momentum of final-state particles $B$ and $C$ in the rest frame of the parent particle $A$ (with $s_i = p_i^2$), and the explicit recursive expansion of the barrier factor $B_L(Q_{abc})$ can be found directly in Ref.~[\onlinecite{Zou2003}]. The projection operator $P^{(L)}(p_A)$ is constructed by summing over all possible spin projections of the spin wave functions:
\begin{widetext}
\begin{equation}
    \sum_{\lambda_B=-J_B}^{J_B} \epsilon^*_{(J_B)\mu_1\cdots\mu_{J_B}}(p_B, \lambda_B) \, \epsilon_{(J_B)\nu_1\cdots\nu_{J_B}}(p_B, \lambda_B) = P^{(J_B)}_{\mu_1\cdots\mu_{J_B}\nu_1\cdots\nu_{J_B}}(p_B)
\end{equation}
\end{widetext}

The fourth term is the final-state spin-coupling tensor $\Omega(S_{BC}; J_B, J_C)$, which describes the coupling of the spin degrees of freedom of daughter particles $B$ and $C$ into the total spin $S_{BC}$. Its general expression is:
\begin{widetext}
\begin{equation}
    \Omega(S_{BC}; J_B, J_C) = P^{(S_{BC})}(p_A) \, \tilde{E}(S_{BC}, J_B, J_C) \, \epsilon^*_{(J_B)}(p_B, \lambda_B) \, \epsilon^*_{(J_C)}(p_C, \lambda_C)
\end{equation}
\end{widetext}
Here, $P^{(S_{BC})}(p_A)$ is the parent-frame spin projection operator. 
During cascade assembly, the sub-amplitude tensors $U_B$ and $U_C$ of secondary particles $B$ and $C$ are constructed within their respective subsystem rest frames, satisfying transversality and tracelessness only with respect to their own four-momenta $p_B$ and $p_C$. 
In the rest frame of the parent particle $A$, however, the secondary particles are in motion. 
If $U_B$ and $U_C$ were contracted directly, the resulting tensor in the reference frame of parent particle $A$ would no longer purely correspond to an irreducible representation of spin $S_{BC}$, but would instead be contaminated by longitudinal components along $p_A$ or non-zero trace components. 
Therefore, it is necessary to introduce $P^{(S_{BC})}(p_A)$ to enforce symmetry, tracelessness, and transversality (the Rarita-Schwinger conditions) of the coupled tensor in the parent particle rest frame. 

In a multi-body cascade decay, if daughter particle $B$ (or $C$) is an unobservable intermediate resonance that undergoes a subsequent two-body decay $B \to D + E$, the amplitude of the secondary decay vertex has the same four-term structure described above:
\begin{equation}
    U_{B\to DE}(\lambda_B) = \epsilon_{(J_B)\nu_1\cdots\nu_{J_B}}(p_B, \lambda_B) \, \mathcal{M}_B^{\nu_1\cdots\nu_{J_B}}(L_{DE}, S_{DE})
\end{equation}
where $\mathcal{M}_B = \tilde{E}(J_B, L_{DE}, S_{DE}) \, T(L_{DE}; p_D, p_E) \, \Omega(S_{DE}; J_D, J_E)$ is the core covariant tensor of the secondary vertex after stripping off external wave functions. 
The conjugate polarization tensor $\epsilon^*_{(J_B)}(p_B, \lambda_B)$ in the fourth term $\Omega(S_{BC})$ of the primary amplitude must be connected to the polarization tensor $\epsilon_{(J_B)}(p_B, \lambda_B)$ in the first term of the secondary amplitude. 
Because the intermediate state is an unobservable particle, one must sum over all possible spin projection directions of intermediate particle $B$, which yields the rank-$J_B$ spin projection operator $P^{(J_B)}(p_B)$ evaluated at its own four-momentum $p_B$.

\subsection{General Form of Single-Photon Radiative Decay Vertices}
Consider a single-photon radiative decay process $A(J_A, P_A; p_A) \to \gamma(q) + X(J_X, P_X; K)$, where $p_A$ is the four-momentum of the parent particle $A$, $q$ is the four-momentum of the emitted real photon, and $K$ is the four-momentum of the recoil hadronic system $X$. 
Unlike massive hadrons, a real photon is a massless spin-1 particle ($q^2 = 0$) that physically possesses only two transverse polarization degrees of freedom with helicities $\lambda_\gamma = \pm 1$, and no longitudinal polarization component. 

The number of independent partial-wave amplitudes $N$ for the single-photon radiative decay $A(J_M) \to \gamma + X(J_X)$ must be strictly identical between the helicity representation and the $L$-$S$ coupling representation [R9]. By helicity conservation and parity conservation, the number of independent radiative amplitudes $N$ is uniquely determined by the parent particle spin $J_M$ and the recoil particle spin $J_X$:
\begin{equation}
    N = \begin{cases} 2\min(J_M, J_X) + 1, & J_M \neq 0,\ J_X \neq 0, J_X \neq J_M \\ 2\min(J_M, J_X), & J_X = J_M \neq 0 \\ 1, & (J_M = 0, J_X \ge 1) \\ & \text{ or } (J_X = 0, J_M \ge 1) \\ 0 \ (\text{forbidden}), & J_M = J_X = 0 \end{cases}
    \label{num_amp_rad}
\end{equation}

Within the covariant $L$-$S$ tensor formalism, the radiative decay vertex can generally be written as: 
\begin{equation}
    U_{A\to\gamma X} = \epsilon^{\mu_1\cdots\mu_{J_M}}(p_A) \, e^\lambda(q) \, \hat{X}^{\nu_1\cdots\nu_{J_X}}(K) \, V_{\mu_1\cdots\mu_{J_M}\lambda\nu_1\cdots\nu_{J_X}}
\end{equation}
where $\epsilon^{\mu_1 \cdots \mu_{J_M}}(p_A)$ is the parent particle wave function, $e^\lambda(q)$ is the photon polarization vector, and $\hat{X}^{\nu_1 \cdots \nu_{J_X}}(K)$ is the purely hadronic decay sub-amplitude.

On this basis, the radiative vertex tensor $V_{\mu_1 \cdots \mu_{J_M}\lambda\nu_1 \cdots \nu_{J_X}}$ is required to strictly satisfy transversality with respect to the photon polarization index $\lambda$. 
To ensure that this transversality condition is identically satisfied when constructing radiative vertices of arbitrary high spin, two rank-2 fundamental transverse tensors are introduced according to the parity nature of the decay vertex:
\begin{enumerate}
    \item Normal-Parity fundamental tensor $K_{\mu\nu}$:
    \begin{equation}
        K_{\mu\nu} = (p_A \cdot q) g_{\mu\nu} - p_{A\nu} q_\mu
    \end{equation}
    \item Anomalous-Parity fundamental tensor $S_{\mu\nu}$:
    \begin{equation}
        S_{\mu\nu} = \epsilon_{\mu\nu\alpha\beta} p_A^\alpha q^\beta
    \end{equation}
\end{enumerate}
Denoting the fundamental transverse tensor as $T_{\alpha\lambda} \in \{K_{\alpha\lambda}, S_{\alpha\lambda}\}$, since $\epsilon(p_A)$ is transverse to $p_A$ and $\hat{X}(K)$ is transverse to $K$, combined with four-momentum conservation $p_A = q + K$, any unpaired parent particle index can only contract with the photon four-momentum $q_\mu$, while any unpaired recoil particle index uniformly contracts with the parent particle four-momentum $p_{A\nu}$. 
Exploiting the exchange symmetry of spin indices, vertices are classified into two categories depending on whether the first index $\alpha$ of $T_{\alpha\lambda}$ contracts with an index of $\hat{X}$ (Type I) or an index of $\epsilon$ (Type II), and universal recursive formulas are obtained by successively replacing the metric tensors $g_{\mu\nu}$ with momentum products $q_\mu p_{A\nu}$:
\begin{enumerate}
    \item Single-sided zero-spin case: There is only 1 independent amplitude. 
    \begin{eqnarray}
        &V_{\mu_1 \cdots \mu_J\lambda} = T_{\mu_1\lambda}q_{\mu_2} \cdots q_{\mu_J}, &J_X = 0\\
        &V_{\lambda\nu_1 \cdots \nu_j} = T_{\nu_1\lambda}p_{A\nu_2} \cdots p_{A\nu_j}, &J_M = 0
    \end{eqnarray}    
    \item Unequal non-zero spin case: Taking $0 < J_M = J < J_X = j$ as an example, there are $2J + 1$ independent amplitudes in total, generated recursively by two sequences: 
    \begin{itemize}
        \item $T_{\nu_j\lambda}$ type, with the number of internal metrics $k = J, J - 1, \dots, 0$, yielding $J + 1$ terms in total: 
        \begin{eqnarray*}
            V^{(1, k)}_{\mu_1\cdots\mu_J\lambda\nu_1\cdots\nu_j} = T_{\nu_j\lambda} \left(\prod_{m=1}^{k} g_{\mu_m\nu_m}\right)\left(\prod_{m=k+1}^{J} q_{\mu_m} p_{A\nu_m}\right) p_{A\nu_{J+1}}\cdots p_{A\nu_{j-1}}
        \end{eqnarray*}
        \item $T_{\mu_J\lambda}$ type, with the number of internal metrics $k = J - 1, J - 2, \dots, 0$, yielding $J$ terms in total: 
        \begin{eqnarray*}
            V^{(2, k)}_{\mu_1\cdots\mu_J\lambda\nu_1\cdots\nu_j} = T_{\mu_J\lambda} \left(\prod_{m=1}^{k} g_{\mu_m\nu_m}\right)\left(\prod_{m=k+1}^{J-1} q_{\mu_m} p_{A\nu_m}\right) p_{A\nu_J} p_{A\nu_{J+1}}\cdots p_{A\nu_j}
        \end{eqnarray*}
    \end{itemize}
    \item Equal non-zero spin case: There are $2J$ independent amplitudes in total, where the number of metric pairings for both types of vertices decreases from $J - 1$ down to $0$ (each contributing $J$ terms):
        \begin{equation}
            V^{(1, k)}_{\mu_1\cdots\mu_J\lambda\nu_1\cdots\nu_J} = T_{\nu_J\lambda} \left(\prod_{m=1}^{k} g_{\mu_m\nu_m}\right) \left(\prod_{m=k+1}^{J-1} q_{\mu_m} p_{A\nu_m}\right) q_{\mu_J}
        \end{equation}
        \begin{equation}
            V^{(2, k)}_{\mu_1\cdots\mu_J\lambda\nu_1\cdots\nu_J} = T_{\mu_J\lambda} \left(\prod_{m=1}^{k} g_{\mu_m\nu_m}\right) \left(\prod_{m=k+1}^{J-1} q_{\mu_m} p_{A\nu_m}\right) p_{A\nu_J}
        \end{equation}
\end{enumerate}
Whenever a metric factor $g_{\mu\nu}$ is replaced by $q_\mu p_{A\nu}$ in the recursion, the number of explicit four-momenta in the vertex increases by 2, corresponding to an increase of 2 in the orbital angular momentum rank. 
Let $N_q$ and $N_p$ denote the total numbers of explicit $q$ and $p_A$ momentum factors appearing in the current term, respectively; then the orbital angular momentum rank $L_{\text{rad}}$ of the matching Blatt-Weisskopf barrier factor $B_{L_{\text{rad}}}(Q_{A\gamma X})$ is determined by the following rules: for a normal-parity vertex ($T = K$), $L_{\text{rad}} = N_q + N_p$; for an anomalous-parity vertex ($T = S$), Type I ($S_{\nu\lambda}$) corresponds to $L_{\text{rad}} = N_q + N_p + 1$, whereas Type II ($S_{\mu\lambda}$) corresponds to $L_{\text{rad}} = \max(0, N_q + N_p - 1)$.

\subsection{Contraction Procedure Across Cascade Vertices}
Let $\mathcal{M}_{i}^{\nu_1 \cdots \nu_{J_i}}(p_i)$ (of rank $J_i$) be the unprojected covariant sub-amplitude tensor of daughter particle $i$ ($i \in \{B, C\}$) recursively passed up from the subsequent decay vertex. 
If daughter particle $i$ is a final-state stable particle (a leaf node of the decay tree), its sub-amplitude reduces to a scalar $\hat{\mathcal{M}}_i = 1$. 
If daughter particle $i$ is an unobservable intermediate resonance with spin $J_i > 0$, summing over all its spin projections is equivalent to applying the spin projection operator $P^{(J_i)}(p_i)$ evaluated at its own four-momentum $p_i$, thereby generating a cascade bridge tensor $\hat{\mathcal{M}}_i(p_i)$ that satisfies the Rarita-Schwinger conditions in the $p_i$ frame:
\begin{equation}
    \hat{\mathcal{M}}_{i}^{\mu_1\cdots\mu_{J_i}}(p_i) = P^{(J_i)\mu_1\cdots\mu_{J_i}\rho_1\cdots\rho_{J_i}}(p_i) \, g_{\rho_1\nu_1}\cdots g_{\rho_{J_i}\nu_{J_i}} \mathcal{M}_{i}^{\nu_1\cdots\nu_{J_i}}(p_i)
\end{equation}
The two bridge-projected subsystem pure-spin tensors $\hat{\mathcal{M}}_B(p_B)$ (of rank $J_B$) and $\hat{\mathcal{M}}_C(p_C)$ (of rank $J_C$) are coupled and contracted toward the target rank $r_{\text{target}} = S_{BC}$ to obtain the preliminary spin-coupling tensor $\tilde{\Omega}^{(S_{BC})}_{\gamma_1 \cdots \gamma_{S_{BC}}}$. 
Subsequently, when the coupled total spin $S_{BC} > 0$, to filter out the longitudinal and parent-frame trace components induced by Lorentz boosts ($p_B, p_C \ne p_A$) in the rest frame of parent particle $A$, the spin projection operator $P^{(S_{BC})}(p_A)$ evaluated at the parent particle four-momentum $p_A$ is further applied, yielding the parent-frame pure-spin coupling tensor $\Omega^{(S_{BC})}(p_A)$:
\begin{equation}
    \Omega^{(S_{BC})}_{\alpha_1\cdots\alpha_{S_{BC}}}(p_A) = P^{(S_{BC})}_{\alpha_1\cdots\alpha_{S_{BC}}\beta_1\cdots\beta_{S_{BC}}}(p_A) \, g^{\beta_1\gamma_1}\cdots g^{\beta_{S_{BC}}\gamma_{S_{BC}}} \tilde{\Omega}^{(S_{BC})}_{\gamma_1\cdots\gamma_{S_{BC}}}
\end{equation}
Next, the rank-$L_{BC}$ covariant orbital angular momentum tensor $\tilde{t}^{(L_{BC})}_{\sigma_1 \cdots \sigma_{L_{BC}}}(p_A, p_B, p_C)$ is constructed using the four-momenta of the current vertex, and contracted (rank $L_{BC}$) with the final-state spin-coupling tensor $\Omega^{(S_{BC})}(p_A)$ (rank $S_{BC}$) toward the target rank $r_{\text{target}} = J_A$. This ultimately generates the covariant amplitude tensor $\mathcal{M}_{A\,\mu_1 \cdots \mu_{J_A}}(p_A)$ for the current parent node (containing $J_A$ free covariant spin indices), which is then passed upward to the preceding decay vertex.

\subsection{Pass-Through Mechanism for Photon Indices in Cascade Decay Amplitudes}
In multi-body cascade decays, a single-photon radiative vertex may reside either at the top-level initial-state particle or within the decay branch of a secondary intermediate resonance. 
If the radiative process occurs at a secondary vertex, the sub-amplitude tensor generated at that vertex must not only carry the spin indices $\mu_1 \cdots \mu_J$ of the current parent particle to participate in hadronic spin couplings at upper-level vertices, but also simultaneously retain the photon polarization index $\lambda$ to be contracted with the external photon polarization vector $e^\lambda(q)$. 

For a secondary radiative vertex, its covariant amplitude tensor is uniformly organized as $\mathcal{M}_{\lambda\mu_1 \cdots \mu_J}$, strictly fixing the photon polarization index $\lambda$ at the first position of the tensor index sequence. 
In all subsequent upper-level vertex operations—including metric contractions $g^{\mu\nu}$, covariant factor contractions $\epsilon_{\alpha\beta\gamma\delta}p^\alpha$, intermediate-state cascade bridge projections $P^{(J_i)}(p_i)$, and parent-frame spin projections $P^{(S)}(p_A)$—the operators act exclusively on and contract within the subspace of the trailing $J$ hadronic spin indices $\{\mu_1 \cdots \mu_J\}$, while transparently passing the leading photon index $\lambda$ upward level by level as an intact spectator free index until it reaches the top-level root node.

\section{Explicit Examples of Covariant Amplitude Construction}
In this section, we select two representative multi-body cascade decay processes in high-energy physics experiments to search for glueballs and exotic hadronic states—namely, the purely hadronic associated decay $J/\psi \to \phi X \ (X \to K^*(892)\bar{K})$ and the single-photon radiative decay $J/\psi \to \gamma X \ (X \to \phi\phi)$—to illustrate the construction framework of covariant tensor amplitudes for intermediate resonances $X$ with various spin-parity quantum numbers $J^{PC}$. 
In the following discussions, the dynamical line shapes parametrizing resonance masses and widths, as well as the complex coupling constants to be fitted, are omitted; only the core covariant amplitude tensors $U$ after stripping off the external polarization wave functions are presented.

\subsection{\texorpdfstring{$J/\psi \to \phi X, \phi \to K^+K^-, X \to K^*(892)\bar{K}, K^*(892) \to K^\pm\pi^\mp$}{ }}
In purely hadronic $J/\psi$ decays, utilizing the associated vector meson $\phi(1020)$ as a flavor filter for strange quark content through the three-step cascade process $J/\psi \to \phi X \ (X \to K^*(892)\bar{K} \to K^\pm \pi^\mp \bar{K})$ provides a systematic approach to investigating the properties of meson resonances and exotic hadronic states with $J^{PC} = 1^{++}, 2^{++}, 2^{-+}$, etc., in the $K^*\bar{K}$ invariant mass spectrum. 
Let the four-momenta of $J/\psi$, $\phi$, $X$, $K^*(892)$, and $\bar{K}$ be denoted by $p_\psi$, $p_\phi$, $p_X$, $p_{K^*}$, and $p_{\bar{K}}$, respectively. 
Both the vector mesons $\phi$ and $K^*(892)$ decay into two pseudoscalar mesons solely via an $(L, S) = (1, 0)$ partial wave; for brevity, their projected decay sub-amplitude tensors are denoted as $T^\alpha(\phi) = P^{\alpha\beta}(p_\phi)t_\beta(p_\phi, p_{K^+}, p_{K^-})$ and $T^\alpha(K^*) = P^{\alpha\beta}(p_{K^*})t_\beta(p_{K^*}, p_{K^\pm}, p_{\pi^\mp})$.

\subsubsection{\texorpdfstring{$J^{PC} = 1^{++}$ for $X$}{}}
For $X$ with $J^{PC} = 1^{++}$, the primary decay ($J/\psi \to \phi X$) allows 3 independent amplitudes denoted as $U_1^{01}, U_1^{21}$, and $U_1^{22}$, while the secondary decay ($X \to K^*(892)\bar{K}$) contains 2 independent amplitudes denoted as $U_2^{01}$ and $U_2^{21}$, yielding a total of 6 independent cascade amplitudes: 
\begin{widetext}
\begin{align}
    U_1^{01}U_2^{01}&=P^{\mu a(1)}(p_{\psi})\varepsilon_{bcda}p_{\psi}^b g^{ce}P_{ef}^{(1)}(p_X)P^{fg(1)}(p_X)g_{gh}T^h(K^*)T^d(\phi),\\
    U_1^{01}U_2^{21}&=P^{\mu a(1)}(p_{\psi})\varepsilon_{bcda}p_{\psi}^b g^{ce}P_{ef}^{(1)}(p_X)g_{gh}t^{gf(2)}(p_X,p_{K^*},p_{K_S^0})P^{hi(1)}(p_X)g_{ij}T^j(K^*)T^d(\phi),\\
    U_1^{21}U_2^{01}&=g_{ab}t^{a\mu(2)}(p_{\psi},p_X,p_\phi)P^{bc(1)}(p_{\psi})\varepsilon_{defc}p_{\psi}^d g^{eg}P_{gh}^{(1)}(p_X)P^{hi(1)}(p_X)g_{ij}T^j(K^*)T^f(\phi),\\
    U_1^{21}U_2^{21}&=g_{ab}t^{a\mu(2)}(p_{\psi},p_X,p_\phi)P^{bc(1)}(p_{\psi})\varepsilon_{defc}p_{\psi}^d g^{eg}P_{gh}^{(1)}(p_X)\notag\\
    &\quad\times g_{ij}t^{ih(2)}(p_X,p_{K^*},p_{K_S^0})P^{jk(1)}(p_X)g_{kl}T^l(K^*)T^f(\phi),\\
    U^{22}_1U_2^{01}&=g^{\mu a}\varepsilon_{bcda}p_{\psi}^b g_{ef}t^{ce(2)}(p_{\psi},p_X,p_\phi)P^{dfgh(2)}(p_{\psi})P_{gi}^{(1)}(p_X)P^{ij(1)}(p_X)g_{jk}T^k(K^*)g_{hl}T^l(\phi),\\
    U^{22}_1U_2^{21}&=g^{\mu a}\varepsilon_{bcda}p_{\psi}^b g_{ef}t^{ce(2)}(p_{\psi},p_X,p_\phi)P^{dfgh(2)}(p_{\psi})P_{gi}^{(1)}(p_X)\notag\\
    &\quad\times g_{jk}t^{ji(2)}(p_X,p_{K^*},p_{K_S^0})P^{kl(1)}(p_X)g_{lm}T^m(K^*)g_{hn}T^n(\phi).
\end{align}
\end{widetext}

\subsubsection{\texorpdfstring{$J^{PC} = 2^{++}$ for $X$}{}}
When $X$ has $J^{PC} = 2^{++}$, the primary decay vertex possesses five independent amplitudes, whereas each of the remaining subsequent decay vertices permits only a single $(L, S)$ configuration ($(L_2, S_2) = (2, 1)$ for $X(2^{++}) \to K^*(892)\bar{K}$). Denoting the overall cascade amplitude corresponding to different orbital angular momentum and spin combinations $(L, S)$ at the primary decay vertex as $U^{LS}$, the five independent amplitudes are written as: 

\begin{widetext}
\begin{align}
    U^{01}&=P^{\mu a(1)}(p_{\psi})P_{bacd}^{(2)}(p_X)g^{ce}\varepsilon_{fghe}p_X^f t^{gd(2)}(p_X,p_{K^*},p_{K_S^0})P^{hi(1)}(p_X)g_{ij}T^j(K^*)T^b(\phi),\\
    U^{21}&=g_{ab}t^{a\mu(2)}(p_{\psi},p_X,p_\phi)P^{bc(1)}(p_{\psi})P_{dcef}^{(2)}(p_X)\notag\\
    &\quad\times g^{eg}\varepsilon_{hijg}p_X^h t^{if(2)}(p_X,p_{K^*},p_{K_S^0})P^{jk(1)}(p_X)g_{kl}T^l(K^*)T^d(\phi),\\
    U^{22}&=g^{\mu a}\varepsilon_{bcda}p_{\psi}^b g_{ef}t^{ce(2)}(p_{\psi},p_X,p_\phi)P^{dfgh(2)}(p_{\psi})\notag\\
    &\quad\times\varepsilon_{ijkg}p_{\psi}^i g^{jl}P_{lhmn}^{(2)}(p_X)g^{mo}\varepsilon_{pqro}p_X^p t^{qn(2)}(p_X,p_{K^*},p_{K_S^0})P^{rs(1)}(p_X)g_{st}T^t(K^*)T^k(\phi),\\
    U^{23}&=g_{ab}g_{cd}t^{ac(2)}(p_{\psi},p_X,p_\phi)P^{bd\mu efg(3)}(p_{\psi})P_{efhi}^{(2)}(p_X)\notag\\
    &\quad\times g^{hj}\varepsilon_{klmj}p_X^k t^{li(2)}(p_X,p_{K^*},p_{K_S^0})P^{mn(1)}(p_X)g_{no}T^o(K^*)g_{gp}T^p(\phi),\\
    U^{43}&=g_{ab}g_{cd}g_{ef}t^{ace\mu(4)}(p_{\psi},p_X,p_\phi)P^{bdfghi(3)}(p_{\psi})P_{ghjk}^{(2)}(p_X)\notag\\
    &\quad\times g^{jl}\varepsilon_{mnol}p_X^m t^{nk(2)}(p_X,p_{K^*},p_{K_S^0})P^{op(1)}(p_X)g_{pq}T^q(K^*)g_{ir}T^r(\phi).
\end{align}
\end{widetext}

\subsubsection{\texorpdfstring{$J^{PC} = 2^{-+}$ for $X$}{}}
For $X$ with $J^{PC} = 2^{-+}$, there are 4 independent amplitudes at the primary decay vertex ($U_1^{11}, U_1^{12}, U_1^{32}, U_1^{33}$) and 2 independent amplitudes at the secondary decay vertex ($U_2^{11}, U_2^{31}$), resulting in a total of 8 independent cascade amplitudes:
\begin{widetext}
\begin{align}
    U_1^{11}U_2^{11}&=g^{\mu a}\varepsilon_{bcda}p_{\psi}^b t^{c(1)}(p_{\psi},p_X,p_\phi)P^{de(1)}(p_{\psi})P_{fegh}^{(2)}(p_X)\\
    &\quad\times t^{g(1)}(p_X,p_{K^*},p_{K_S^0})P^{hi(1)}(p_X)g_{ij}T^j(K^*)T^f(\phi),\\
    U_1^{11}U_2^{31}&=g^{\mu a}\varepsilon_{bcda}p_{\psi}^b t^{c(1)}(p_{\psi},p_X,p_\phi)P^{de(1)}(p_{\psi})P_{fegh}^{(2)}(p_X)\\
    &\quad\times g_{ij}t^{igh(3)}(p_X,p_{K^*},p_{K_S^0})P^{jk(1)}(p_X)g_{kl}T^l(K^*)T^f(\phi),\\
    U_1^{12}U_2^{11}&=g_{ab}t^{a(1)}(p_{\psi},p_X,p_\phi)P^{b\mu cd(2)}(p_{\psi})\varepsilon_{efgc}p_{\psi}^e g^{fh}P_{hdij}^{(2)}(p_X)\\
    &\quad\times t^{i(1)}(p_X,p_{K^*},p_{K_S^0})P^{jk(1)}(p_X)g_{kl}T^l(K^*)T^g(\phi),\\
    U_1^{12}U_2^{31}&=g_{ab}t^{a(1)}(p_{\psi},p_X,p_\phi)P^{b\mu cd(2)}(p_{\psi})\varepsilon_{efgc}p_{\psi}^e g^{fh}P_{hdij}^{(2)}(p_X)\notag\\
    &\quad\times g_{kl}t^{kij(3)}(p_X,p_{K^*},p_{K_S^0})P^{lm(1)}(p_X)g_{mn}T^n(K^*)T^g(\phi),\\
    U_1^{32}U_2^{11}&=g_{ab}g_{cd}t^{ac\mu(3)}(p_{\psi},p_X,p_\phi)P^{bdef(2)}(p_{\psi})\varepsilon_{ghie}p_{\psi}^g g^{hj}P_{jfkl}^{(2)}(p_X)\notag\\
    &\quad\times t^{k(1)}(p_X,p_{K^*},p_{K_S^0})P^{lm(1)}(p_X)g_{mn}T^n(K^*)T^i(\phi),\\
    U_1^{32}U_2^{31}&=g_{ab}g_{cd}t^{ac\mu(3)}(p_{\psi},p_X,p_\phi)P^{bdef(2)}(p_{\psi})\varepsilon_{ghie}p_{\psi}^g g^{hj}P_{jfkl}^{(2)}(p_X)\notag\\
    &\quad\times g_{mn}t^{mkl(3)}(p_X,p_{K^*},p_{K_S^0})P^{no(1)}(p_X)g_{op}T^p(K^*)T^i(\phi),\\
    U_1^{33}U_2^{11}&=g^{\mu a}\varepsilon_{bcda}p_{\psi}^b g_{ef}g_{gh}t^{ceg(3)}(p_{\psi},p_X,p_\phi)P^{dfhijk(3)}(p_{\psi})P_{ijlm}^{(2)}(p_X)\notag\\
    &\quad\times t^{l(1)}(p_X,p_{K^*},p_{K_S^0})P^{mn(1)}(p_X)g_{no}T^o(K^*)g_{kp}T^p(\phi),\\
    U_1^{33}U_2^{31}&=g^{\mu a}\varepsilon_{bcda}p_{\psi}^b g_{ef}g_{gh}t^{ceg(3)}(p_{\psi},p_X,p_\phi)P^{dfhijk(3)}(p_{\psi})P_{ijlm}^{(2)}(p_X)\notag\\
    &\quad\times g_{no}t^{nlm(3)}(p_X,p_{K^*},p_{K_S^0})P^{op(1)}(p_X)g_{pq}T^q(K^*)g_{kr}T^r(\phi).
\end{align}
\end{widetext}

\subsection{\texorpdfstring{$J/\psi \to \gamma X, X \to \phi\phi, \phi \to K^+K^-$}{}}
The radiative decay of $J/\psi$ into a double-vector-meson final state, $J/\psi \to \gamma \phi \phi \to \gamma K^+ K^- K^+ K^-$, is a quintessential gluon-rich process and serves as a prime hunting ground in experiments for tensor glueballs ($J^{PC} = 2^{++}$) and pseudotensor exotic states or glueball candidates ($J^{PC} = 2^{-+}$). 
Let the four-momentum of the initial-state $J/\psi$ be $p_\psi$, that of the radiative photon be $q$, that of the intermediate resonance $X$ be $p_X = p_{\phi_1} + p_{\phi_2}$, and those of the two $\phi$ mesons be $p_{\phi_1} = p_{K^+_1} + p_{K^-_1}$ and $p_{\phi_2} = p_{K^+_2} + p_{K^-_2}$, respectively.

\subsubsection{\texorpdfstring{$J^{PC} = 2^{++}$ for $X$}{}}
According to the counting and recursive formulas derived in Eq.~(\ref{num_amp_rad}), the primary decay vertex ($J/\psi \to \gamma X$) possesses 3 independent amplitudes, while the secondary decay vertex ($X \to \phi\phi$) has 5 independent amplitudes, resulting in a total of 15 independent partial-wave amplitudes. The primary vertex is a normal-parity radiative vertex, whose three independent structures are denoted as $U_1, U_2$, and $U_3$, corresponding to:
\begin{eqnarray}
    U_1&=T^{a\lambda}g^{\mu b}X_{ab},\\
    U_2&=T^{a\lambda}q^\mu p^bX_{ab},\\
    U_3&=T^{\mu\lambda}p^a p^bX_{ab},
\end{eqnarray}
where $X_{ab}$ represents the abstract sub-amplitude tensor of the secondary decay, and the transverse fundamental tensor $T$ takes the normal-parity form $K$. 
The secondary decay amplitudes, constructed via orbital-spin ($L$-$S$) angular momentum coupling, are denoted as $A^{LS}$, where $L$ and $S$ represent the relative orbital angular momentum and coupled total spin of the secondary $\phi\phi$ system, respectively. Using the shorthand notation $B_L(p_\psi, q, p_X) = B_L$, all 15 independent cascade amplitudes are explicitly constructed as follows: 
\begin{widetext}
\begin{align}
    U_1A^{02}&=T^{a\lambda}g^{\mu b}P_{abcd}^{(2)}(p_X)t^{(0)}(p_X,p_{\phi_1},p_{\phi_2})P^{cdef(2)}(p_X)g_{eg}g_{fh}T^g(\phi_1)T^h(\phi_2),\\
    U_1A^{20}&=T^{a\lambda}g^{\mu b}P_{abcd}^{(2)}(p_X)t^{cd(2)}(p_X,p_{\phi_1},p_{\phi_2})g_{ef}T^e(\phi_1)T^f(\phi_2),\\
    U_1A^{21}&=T^{a\lambda}g^{\mu b}P_{abcd}^{(2)}(p_X)g^{ce}\varepsilon_{fghe}p_X^f t^{gd(2)}(p_X,p_{\phi_1},p_{\phi_2})P^{hi(1)}(p_X)\varepsilon_{jkli}p_X^j T^k(\phi_1)T^l(\phi_2),\\
    U_1A^{22}&=T^{a\lambda}g^{\mu b}P_{abcd}^{(2)}(p_X)g_{ef}t^{ec(2)}(p_X,p_{\phi_1},p_{\phi_2})P^{fdgh(2)}(p_X)g_{gi}g_{hj}T^i(\phi_1)T^j(\phi_2),\\
    U_1A^{42}&=T^{a\lambda}g^{\mu b}P_{abcd}^{(2)}(p_X)g_{ef}g_{gh}t^{egcd(4)}(p_X,p_{\phi_1},p_{\phi_2})P^{fhij(2)}(p_X)g_{ik}g_{jl}T^k(\phi_1)T^l(\phi_2),\\
    U_2A^{20}&=B_2T^{a\lambda}q^\mu p_{\psi}^b P_{abcd}^{(2)}(p_X)t^{cd(2)}(p_X,p_{\phi_1},p_{\phi_2})g_{ef}T^e(\phi_1)T^f(\phi_2),\\
    U_2A^{21}&=B_2T^{a\lambda}q^\mu p_{\psi}^b P_{abcd}^{(2)}(p_X)g^{ce}\varepsilon_{fghe}p_X^f t^{gd(2)}(p_X,p_{\phi_1},p_{\phi_2})P^{hi(1)}(p_X)\varepsilon_{jkli}p_X^j T^k(\phi_1)T^l(\phi_2),\\
    U_2A^{02}&=B_2T^{a\lambda}q^\mu p_{\psi}^b P_{abcd}^{(2)}(p_X)t^{(0)}(p_X,p_{\phi_1},p_{\phi_2})P^{cdef(2)}(p_X)g_{eg}g_{fh}T^g(\phi_1)T^h(\phi_2),\\
    U_2A^{22}&=B_2T^{a\lambda}q^\mu p_{\psi}^b P_{abcd}^{(2)}(p_X)g_{ef}t^{ec(2)}(p_X,p_{\phi_1},p_{\phi_2})P^{fdgh(2)}(p_X)g_{gi}g_{hj}T^i(\phi_1)T^j(\phi_2),\\
    U_2A^{42}&=B_2T^{a\lambda}q^\mu p_{\psi}^b P_{abcd}^{(2)}(p_X)g_{ef}g_{gh}t^{egcd(4)}(p_X,p_{\phi_1},p_{\phi_2})P^{fhij(2)}(p_X)g_{ik}g_{jl}T^k(\phi_1)T^l(\phi_2),\\
    U_3A^{02}&=B_2T^{\mu\lambda}p_{\psi}^a p_{\psi}^b P_{abcd}^{(2)}(p_X)t^{(0)}(p_X,p_{\phi_1},p_{\phi_2})P^{cdef(2)}(p_X)g_{eg}g_{fh}T^g(\phi_1)T^h(\phi_2),\\
    U_3A^{20}&=B_2T^{\mu\lambda}p_{\psi}^a p_{\psi}^b P_{abcd}^{(2)}(p_X)t^{cd(2)}(p_X,p_{\phi_1},p_{\phi_2})g_{ef}T^e(\phi_1)T^f(\phi_2),\\
    U_3A^{21}&=B_2T^{\mu\lambda}p_{\psi}^a p_{\psi}^b P_{abcd}^{(2)}(p_X)g^{ce}\varepsilon_{fghe}p_X^f t^{gd(2)}(p_X,p_{\phi_1},p_{\phi_2})P^{hi(1)}(p_X)\varepsilon_{jkli}p_X^j T^k(\phi_1)T^l(\phi_2),\\
    U_3A^{22}&=B_2T^{\mu\lambda}p_{\psi}^a p_{\psi}^b P_{abcd}^{(2)}(p_X)g_{ef}t^{ec(2)}(p_X,p_{\phi_1},p_{\phi_2})P^{fdgh(2)}(p_X)g_{gi}g_{hj}T^i(\phi_1)T^j(\phi_2),\\
    U_3A^{42}&=B_2T^{\mu\lambda}p_{\psi}^a p_{\psi}^b P_{abcd}^{(2)}(p_X)g_{ef}g_{gh}t^{egcd(4)}(p_X,p_{\phi_1},p_{\phi_2})P^{fhij(2)}(p_X)g_{ik}g_{jl}T^k(\phi_1)T^l(\phi_2).
\end{align}
\end{widetext}

\subsubsection{\texorpdfstring{$J^{PC} = 2^{-+}$ for $X$}{}}
In this case, the primary decay vertex features three anomalous-parity radiative structures, denoted as $V_1, V_2$, and $V_3$, whose formal expressions are identical to those in the $2^{++}$ case except that the fundamental transverse tensor $T$ is replaced by the anomalous-parity tensor $S$. 
Combined with the four independent amplitudes allowed in the secondary decay, there are a total of 12 independent partial-wave amplitudes, which are given respectively by 
\begin{widetext}
\begin{align}
    V_1A^{11}&=B_1T^{a\lambda}g^{\mu b}P_{abcd}^{(2)}(p_X)t^{c(1)}(p_X,p_{\phi_1},p_{\phi_2})P^{de(1)}(p_X)\varepsilon_{fghe}p_X^f T^g(\phi_1)T^h(\phi_2),\\
    V_1A^{12}&=B_1T^{a\lambda}g^{\mu b}P_{abcd}^{(2)}(p_X)g^{ce}\varepsilon_{fghe}p_X^f t^{g(1)}(p_X,p_{\phi_1},p_{\phi_2})P^{hdij(2)}(p_X)g_{ik}g_{jl}T^k(\phi_1)T^l(\phi_2),\\
    V_1A^{31}&=B_1T^{a\lambda}g^{\mu b}P_{abcd}^{(2)}(p_X)g_{ef}t^{ecd(3)}(p_X,p_{\phi_1},p_{\phi_2})P^{fg(1)}(p_X)\varepsilon_{hijg}p_X^h T^i(\phi_1)T^j(\phi_2),\\
    V_1A^{32}&=B_1T^{a\lambda}g^{\mu b}P_{abcd}^{(2)}(p_X)g^{ce}\varepsilon_{fghe}p_X^f g_{ij}t^{gid(3)}(p_X,p_{\phi_1},p_{\phi_2})P^{hjkl(2)}(p_X)g_{km}g_{ln}T^m(\phi_1)T^n(\phi_2),\\
    V_2A^{11}&=B_3T^{a\lambda}q^\mu p_{\psi}^b P_{abcd}^{(2)}(p_X)t^{c(1)}(p_X,p_{\phi_1},p_{\phi_2})P^{de(1)}(p_X)\varepsilon_{fghe}p_X^f T^g(\phi_1)T^h(\phi_2),\\
    V_2A^{12}&=B_3T^{a\lambda}q^\mu p_{\psi}^b P_{abcd}^{(2)}(p_X)g^{ce}\varepsilon_{fghe}p_X^f t^{g(1)}(p_X,p_{\phi_1},p_{\phi_2})P^{hdij(2)}(p_X)g_{ik}g_{jl}T^k(\phi_1)T^l(\phi_2),\\
    V_2A^{31}&=B_3T^{a\lambda}q^\mu p_{\psi}^b P_{abcd}^{(2)}(p_X)g_{ef}t^{ecd(3)}(p_X,p_{\phi_1},p_{\phi_2})P^{fg(1)}(p_X)\varepsilon_{hijg}p_X^h T^i(\phi_1)T^j(\phi_2),\\
    V_2A^{32}&=B_3T^{a\lambda}q^\mu p_{\psi}^b P_{abcd}^{(2)}(p_X)g^{ce}\varepsilon_{fghe}p_X^f g_{ij}t^{gid(3)}(p_X,p_{\phi_1},p_{\phi_2})P^{hjkl(2)}(p_X)g_{km}g_{ln}T^m(\phi_1)T^n(\phi_2),\\
    V_3A^{11}&= B_1T^{\mu\lambda}p_{\psi}^a p_{\psi}^b P_{abcd}^{(2)}(p_X)t^{c(1)}(p_X,p_{\phi_1},p_{\phi_2})P^{de(1)}(p_X)\varepsilon_{fghe}p_X^f T^g(\phi_1)T^h(\phi_2),\\
    V_3A^{12}&=B_1T^{\mu\lambda}p_{\psi}^a p_{\psi}^b P_{abcd}^{(2)}(p_X)g^{ce}\varepsilon_{fghe}p_X^f t^{g(1)}(p_X,p_{\phi_1},p_{\phi_2})P^{hdij(2)}(p_X)g_{ik}g_{jl}T^k(\phi_1)T^l(\phi_2),\\
    V_3A^{31}&=B_1T^{\mu\lambda}p_{\psi}^a p_{\psi}^b P_{abcd}^{(2)}(p_X)g_{ef}t^{ecd(3)}(p_X,p_{\phi_1},p_{\phi_2})P^{fg(1)}(p_X)\varepsilon_{hijg}p_X^h T^i(\phi_1)T^j(\phi_2),\\
    V_3A^{32}&=B_1T^{\mu\lambda}p_{\psi}^a p_{\psi}^b P_{abcd}^{(2)}(p_X)g^{ce}\varepsilon_{fghe}p_X^f g_{ij}t^{gid(3)}(p_X,p_{\phi_1},p_{\phi_2})P^{hjkl(2)}(p_X)g_{km}g_{ln}T^m(\phi_1)T^n(\phi_2).
\end{align}
\end{widetext}

\section{Summary}

First, based on the decoupled structure of the total partial-wave amplitude into external polarization wave functions, complex coupling parameters, and the core covariant amplitude tensor $U$, this paper establishes a universal covariant $L$-$S$ tensor construction method for multi-body purely hadronic cascade decays. By explicitly introducing the cascade bridge projection operator $P^{(J_i)}(p_i)$—which corresponds to the polarization sum of intermediate resonances—along with the definite-spin projection operator $P^{(S_{BC})}(p_A)$ evaluated in the parent particle rest frame, this approach strictly guarantees the symmetry, tracelessness, and transversality (the Rarita-Schwinger conditions) of tensors at each level when assembled across different reference frames, thereby fundamentally eliminating the unphysical longitudinal and trace contaminations induced in the parent frame by Lorentz boosts of secondary particles. Second, we propose a general covariant tensor formalism for single-photon radiative decays $A(J_M) \to \gamma + X(J_X)$ that strictly satisfies the photon transversality condition ($q^\lambda V_{\dots\lambda\dots} = 0$) in the Lorenz gauge. Built upon the normal-parity fundamental tensor $K_{\mu\nu}$ and the anomalous-parity fundamental tensor $S_{\mu\nu}$, dynamic recursive formulas are formulated to construct all independent radiative amplitudes via the systematic replacement $g_{\mu\nu} \to q_\mu p_{A\nu}$, accompanied by precise selection rules for determining the orbital angular momentum rank $L_{\text{rad}}$ of the matching Blatt-Weisskopf centrifugal barrier factor $B_{L_{\text{rad}}}$. Finally, by combining the standardized contraction procedure across cascade vertices with the spectator pass-through mechanism for the dedicated photon polarization index $\lambda$, we establish a complete and systematic framework for constructing covariant amplitude tensors across arbitrary cascade decay topologies.

Unlike traditional literature, which typically constructs tensor expressions only for specific decay channels involving low-spin parent particles (e.g., $J \le 1$) or intermediate resonances with $J \le 2$, the unified recursive formalism developed in this work overcomes the limitations on spin ranks and can be universally extrapolated to parent particles and intermediate resonances of arbitrary high spins $(J_M, J_X)$. The general theoretical formulas presented herein are directly applicable to partial wave analyses of the massive, high-statistics hadronic and radiative cascade decay datasets collected by experiments such as BESIII, LHCb, Belle II, and GlueX. Furthermore, they can be seamlessly integrated into high-performance fitting frameworks that utilize pre-computation and static GPU memory caching (such as PWACG), enabling the efficient and precise extraction of spin-parity quantum numbers and dynamical properties for exotic hadronic states—including glueballs, hybrids, and multiquarks—as well as conventional excited resonances within complex multi-body interfering final states.



\nocite{*}

\bibliography{apssamp}

\end{document}